\documentclass{elsart}
\usepackage{epsfig}
\usepackage{amssymb}

\begin{document}
\begin{frontmatter}
\title{Multichannel quantum defect theory: a quantum Poincar\'e map.}
\author[UNAM,CIC]{F. Leyvraz}, 
\author[CIC]{R. A. M\'endez-S\'anchez\thanksref{Essen}}, 
\author[Spectro]{M. Lombardi\thanksref{Corresponding}} \and
\author[UNAM,CIC]{T. H. Seligman}
\address[UNAM]{Centro de Ciencias F\'{\i}sicas University of Mexico 
(UNAM), Cuernavaca, Mexico}
\address[CIC]{Centro Internacional de Ciencias, Cuernavaca, Mexico}
\address[Spectro]{Laboratoire de Spectrom\'etrie Physique, CNRS and Universit\'e
Joseph-Fourier-Grenoble (UMR 5588), 
BP87, F-38402~Saint-Martin-d'H\`eres~C\'edex, France}
\thanks[Essen]{Present address:
Fachbereich 7, Physik, Universit\"at G.~H. Essen, 45117 Essen}
\thanks[Corresponding]{Corresponding author; Fax: +33~476~514~544;
e-mail: Maurice.Lombardi@ujf-grenoble.fr}

\begin{abstract}
The multichannel quantum defect theory (MQDT) can be reinterpreted
as a quantum Poincar\'e map in representation of angular momentum. 
This has two important implications: we have a paradigm of a 
true quantum Poincar\'e map without semi-classical input and  
we get an entirely new insight into the significance of MQDT.

PACS: 05.45.Mt; 33.80.Rv; 03.65.Sq
\end{abstract}

\end{frontmatter}

In recent years there has been a rapidly growing interest in the quantum 
Poincar\'e map (QPM) 
\cite{Bogo,%
%Bogo2,Bogo3,%
Doron&Smilansky,Dietz&Smilansky,Prosen,Szeredietal,%
%Szeredietal2,%
Gutzwiller,Goodings&Whelan,Haggerty,Prange&Georgeot,Prange}, 
i.e. the quantization of a classical Poincar\'e map, 
for a time independent Hamiltonian system. Bogomolny\cite{Bogo}
started out with a semi-classical formulation.
Among other things he shows that unitarity of the representation is 
reached in the limit $\hbar \to 0$, while this is only approximate
for finite $\hbar$ \cite{Haggerty}. Prosen \cite{Prosen} gives an 
elegant general solution to the unitarity problem at the expense of 
obtaining an infinite matrix for the QPM. 
The semi-classical approach common to most discussions 
causes a number of problems that make the use of this new and powerful 
tool a little obscure.
With other words, the quantum Poincar\'e section implicitly defined by 
Bogomolny, lacks a paradigmatic example where a quantum treatment can be 
performed properly throughout and leads to a finite unitary matrix. 

Multichannel quantum defect theory (MQDT)
\cite{Seaton,Fano,Fano&Rau,Greeneetal,Bordasetal}
and its classical limit \cite{Lombardietal,Lombardi&Seligman,Dietzetal} 
will be shown to provide
the framework for such a paradigm. Indeed we shall see that 
a simplified model of the Rydberg molecule allows to construct a 
classical Poincar\'e map on the unit sphere, whose exact quantization is 
provided by MQDT. Thus the result is necessarily entirely quantal,
exactly unitary and for finite $\hbar$ given in terms of a finite matrix.
We shall show that the results commonly
derived for MQDT are directly properties of the unitary representation 
of this classical map as obtained by MQDT.

After a short description of the model for a Rydberg molecule and the 
simplification introduced in Ref.~\cite{Lombardietal}, 
we proceed to give the quantum map for this case explicitly. 
We illustrate the two important aspects of our result by two 
applications.  First the new interpretation allows modifications of
the MQDT method, that prove particularly effective in near integrable 
systems. Second we proceed to show by way of examples that 
the properties of this map are relevant to the study of chaos and order
in this system.  

Simplifying to the most basic case, these molecules can be viewed as a
rotating system with positive charge and cylindrical symmetry that 
binds one electron in an orbit that is at large distances hydrogenic. 
The classical limit of the MQDT is the following classical model
\cite{Lombardietal}: The motion is
composed of two consecutive steps. (i) when the electron is far
from the molecular core (i.e. most of the time for a Rydberg electron)
it feels only the Coulomb part ($-1/r\/$) of the potential. 
Its orbit is hydrogenic and its angular momentum ${\bf L}$ is fixed
in the laboratory reference frame. 
Meanwhile the core rotates freely with an angular momentum
${\bf N}$ which is also fixed in the laboratory frame. The total angular
momentum ${\bf J} = {\bf L} + {\bf N}$ is always conserved. In
the molecular reference frame, the $OZ$ axis is the cylindrical symmetry
axis of the core.
The core angular momentum ${\bf N}$ points in 
a perpendicular direction, taken as the $OX$ axis. The angles
$\theta_e$ and $\phi_e$ are the polar and azimuthal angles respectively
of the electronic angular momentum ${\bf L}$ in this frame.
During this step, ${\bf L}$
rotates freely around the $OX$ axis. 
(ii) during the so called "collision" step, the
electron senses also the cylindrically symmetric short range part of
the potential of the core. Aside from the energy and ${\bf J}$,  
the projection of ${\bf L}$ onto the core axis
$\Lambda =L\cos \theta _e$ is conserved due to the cylindrical symmetry 
of the core. 
We will add an extra, simplifying, hypothesis, namely that the
magnitude $L$ of ${\bf L}$ remains constant\cite{Lombardietal}. 
This is justified for Rydberg Molecules at least for small $L$'s, but 
the classical and quantum map with this approximation exist for all $L$. 
Thus the collision can be described by a $ \theta _e$--dependent
rotation of ${\bf L}$ around the core axis. The simplest form of
this rotation compatible with the symmetry is \cite{Lombardietal}:
$\delta \varphi _e=K\cos \theta _e$, where $K$ is a coupling constant. 
This simplification is not essential. Notice further that the 
conservation of the total angular momentum ${\bf J}$ implies that the
molecular core feels a simultaneous recoil which changes the direction
and magnitude of ${\bf N}$. This change of $N$ in turn entails a change
of the rotational energy $E_N$ of the core and because of 
conservation of total energy a change of the energy $E_e$ of the electron.
This exchange of energy makes this model much richer than the kicked
spin model \cite{Nakamuraetal}
(which is its limit when $L\ll J$, where this recoil can be neglected).
In particular the energy of the electron may become positive after the 
collision, allowing to treat on equal footing bound and unbound
(ionized) states. Possibility of chaotic motion comes from the conflict
of these two steps, which consist of two rotations around distinct axes
with different laws. The classically chaotic case can be obtained by 
increasing the coupling $K$. Near-integrable cases can be obtained for
small coupling or at resonance, i.e. when the period of the electron is
a multiple of half the period of rotation of the core.

The quantum problem is solved by using the MQDT. The configuration space is
divided by a sphere of radius $r_0$ in a collision ($r<r_0$) and an
asymptotic region ($r>r_0$) for the motion of the electron. $r_0$ is of
the order of the core size and is chosen such that in the asymptotic region
the potential acting on the electron is only Coulombic, whereas in the
collision region it feels both Coulomb and cylindrical potential. The wave
functions for both regions are joined appropriately at $r=r_0$.
The conflict between the two motions is expressed in quantum mechanics
by the existence for the wave function of two bases with different
good quantum numbers (in addition to $J, J_z, L$ and total energy $E$). 
At short distance the Born Oppenheimer
basis is appropriate. The Rydberg electron is strongly bound to the core,
thus quantized in the molecular reference frame
and the additional good quantum number is $\Lambda$. At long distances
the collision basis is appropriate. Here the electron
is uncoupled from the core and the angular momentum $N$
of the core remains a good quantum number.
The collision is described by  phase shifts $\mu_\Lambda$,
which are identical to collision phase shifts if the total energy
is positive enough for all channels to be open, and which are related
to the classical $K$ parameter by
$\mu_\Lambda = \mu_0 -(K / 4\pi L) \Lambda^2$.

We focus on the completely bound situation, when
total energy is low enough for the electron energy to be always negative
whatever the value of $N$ within the allowed range $[ J-L, J+L] $.
Demanding that the electron wave function goes to zero when $r \to \infty$
leads to demanding that the following determinant vanishes
\cite{Bordasetal}, {\it i.e. }
\begin{equation}
\det{\sf S}=\det \left\{U_{N\Lambda} \sin(\pi(\mu_\Lambda+\nu_N))\right\}=0,
\label{eq:S}
\end{equation}
where the {\em unitary} $U$ matrix given by
\begin{equation}
U_{N,\Lambda}=<L, -\Lambda,J,\Lambda|N,0>(-1)^{J-N+\Lambda}
(2-\delta_\Lambda,0)^{1/2}.
\end{equation}
relates the two conflicting bases through a Clebsh Gordan coefficient.
The principal (non integer) action $\nu_N(E)$ 
 of the Coulomb electron,
is related to the electron energy  through
$E=E_N+E_e=N(N+1)/(2I)-1/(2\nu_N^2)$, 
where $I$ is the moment of inertia of the core,
and we use atomic units ($e=m=\hbar=1\/$).
Corresponding wave functions are the eigenkets of $\sf S$
for the eigenvalue zero, labeled by $\Lambda$ in the 
Born Oppenheimer basis: 
\begin{equation}
{\sf S}|A_\Lambda\rangle =0.
\end{equation}
Similarly the eigenbras $\langle B_N|$ of $\sf S$ (eigenkets of ${\sf S}^t$),
labeled by $N$ describe the corresponding wave function in the collision basis.

To proceed notice first that ${\sf S}$ is the imaginary part of
a complex {\em unitary} matrix
\begin{equation}
{\sf E}={\sf C}+ i {\sf S}= 
\left\{U_{N\Lambda} \exp(i\pi(\mu_\Lambda+\nu_N))\right\}.
\label{eq:E}
\end{equation}
This non symmetrical matrix maps the $N$ basis on the $\Lambda$ basis:
it is {\em half} the QPM we look for, and describes the motion
between apogee and perigee. From it we can construct two QPM,
which, by construction, turn out to be 
{\em symmetrical unitary} complex matrices.

${\sf E}^t{\sf E}$ operates in Born Oppenheimer $\Lambda$ space,
and is the exact quantization of a classical Poincar\'e map.
The latter is nearly the classical map used in 
refs.~\cite{Lombardietal,Lombardi&Seligman}. That map on
the unit sphere described the position of ${\bf L}$
in the molecular frame immediately after the collision, 
whereas the present map describes the position of ${\bf L}$
in the middle of the collision (perigee).
That this matrix is the $T$ matrix
defined by Bogomolny\cite{Bogo} to quantize a Poincar\'e map
will be shown by proving that the eigenvalues and eigenfunctions
of the entire system result from 
\begin{equation}
\det(1-T(E_n))=0,
\label{eq:T}
\end{equation} 
i.e. Bogomolny's equation for the quantized energy $E_n$.
Indeed at quantized energies given by Eq.~(\ref{eq:S})
\begin{equation}
{\sf E}^t{\sf E} |A_\Lambda\rangle= 1 |A_\Lambda\rangle,
\end{equation} 
which implies (\ref{eq:T}).
To prove this key point first notice that unitarity
of ${\sf E}$, namely ${\sf E}^{\dag}{\sf E}=
({\sf C}^t-i{\sf S}^t)({\sf C}+i{\sf S})=\mathbb{I}$,
leads to 
${\sf C}^t{\sf C}+{\sf S}^t{\sf S}=\mathbb{I}$ and
${\sf C}^t{\sf S}={\sf S}^t{\sf C}$.
Then ${\sf E}^t{\sf E}=
({\sf C}^t+i{\sf S}^t)({\sf C}+i{\sf S})=
\mathbb{I}-2 {\sf S}^t{\sf S}+ 2 i {\sf C}^t{\sf S}$,
so that if ${\sf S}|A_\Lambda\rangle =0$ then
${\sf E}^t{\sf E}|A_\Lambda\rangle =\mathbb{I}|A_\Lambda\rangle $.

Conversely ${\sf E}{\sf E}^t$ operates in collision
$N$ space and corresponds to a classical map between apogee and apogee
in the laboratory frame. 

We will now compare the traditional way of solving MQDT
to the one implied by our QPM.
The traditional way is to look for zeros of the 
determinant of the non-symmetric matrix ${\sf S}$
of Eq.~(\ref{eq:S}), all of whose elements depend 
in a complex way on energy through $\nu_N(E)$. 
It is computed by a LU or a SVD method 
followed by a root searching algorithm
\cite{NumericalRecipes,Lombardietal}. The present
method is to look for eigenphases of ${\sf E}^t{\sf E}$
or ${\sf E}{\sf E}^t$. They are computed by the diagonalization
of a symmetrical unitary  matrix. This is  
efficient and unproblematic because it diagonalizes
in an orthonormal basis. Finally we search the
zeros of the resulting eigenphases. 

\begin{figure}[htb]
\centerline{\epsfig{file=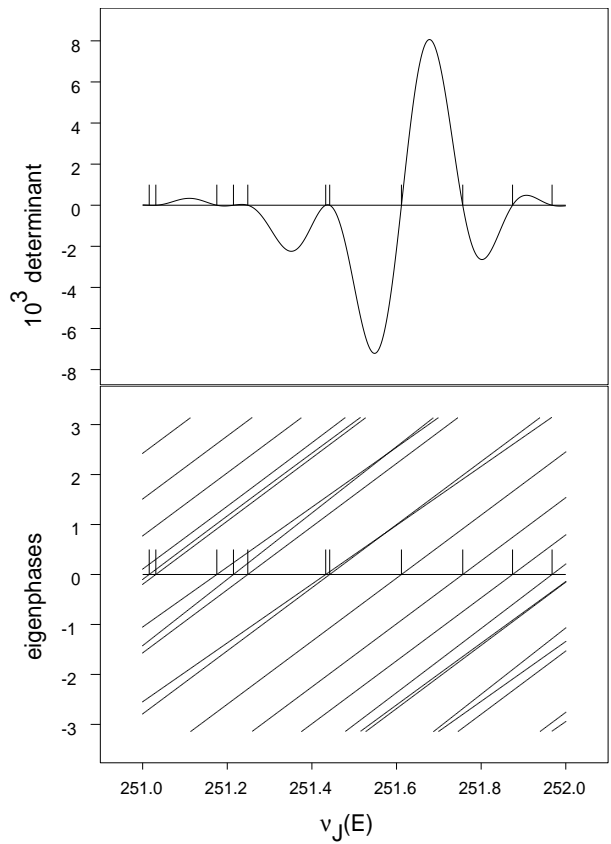,width=8cm}}
\caption{Comparison of Determinant and Eigenphases methods
to solve the MQDT problem. 
Abscissa axis is energy on a scale $\nu_N(E)$ for the
middle $N=J$ value.
}
\label{fig:Integrable}
\end{figure}

The situation is sketched in Fig.~\ref{fig:Integrable}. The 
search for zeros of eigenphases which vary nearly
linearly with energy is obviously much simpler than
the search of zeros of a determinant which is sometimes nearly
tangent to the horizontal axis. Moreover, in this case
of near tangency we had problems to converge the
wave functions because the eigenfunction switches between
two nearly orthogonal values in a very narrow energy range
\cite{Lombardi&Seligman}, requiring the use of the  
more efficient but slower SVD algorithm. 
Such a situation occurs frequently in nearly integrable cases, 
due to the lack of level repulsion.
The diagonalization, on the contrary, gives always correctly
the orthogonal eigenfunctions even if eigenvalues are very close.

\begin{figure}[htb]
\centerline{\epsfig{file=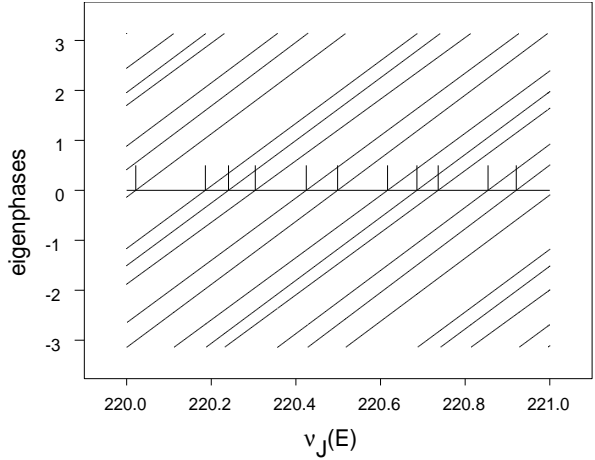,width=8cm}}
\caption{Eigenphases in a chaotic case.}
\label{fig:Chaotique}
\end{figure}

\begin{figure}[htb]
\centerline{\epsfig{file=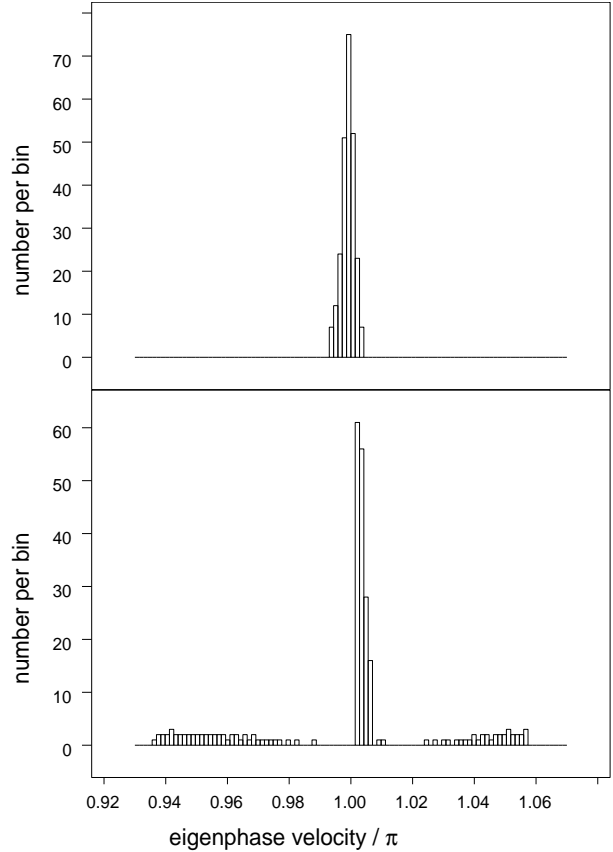,width=8cm}}
\caption{Histograms of the velocities of the eigenphases curves
for a given energy.
Top: chaotic case, bottom: integrable case. These histograms
correspond to the same classical parameters as in 
Figs.~\protect\ref{fig:Integrable} and 
\protect\ref{fig:Chaotique}, 
but $\hbar$ has been divided by 25
to increase statistics}
\label{fig:Histo}
\end{figure}

The implications of this procedure
for the study of quantum chaos is of great interest.
Figure~\ref{fig:Chaotique} displays a chaotic situation.
Comparing eigenphases in the near integrable (Fig.~\ref{fig:Integrable})
and chaotic (Fig.~\ref{fig:Chaotique}) situations, we see that in the 
first case the phases as a function of energy display avoided 
crossings of straight lines running at different angles while for the 
second case these lines are practically parallel.
This shows that the drift of eigenphases as a function of energy displays 
presence or absence of level repulsion or spectral rigidity
more obviously than the energy levels themselves.
This consideration is important in relation
with the theory put forward by two of us
\cite{Leyvraz&Seligman%
%,Leyvraz&Seligman2,Leyvraz&Seligman3%
}, which relates Random Matrix Properties
of eigenvalues of a quantum system to properties of invariance 
under canonical transformations of the structure of the corresponding
classical system (structural invariance). This theory
was developed for maps such as the scattering map, the stroboscopic 
map or the Poincar\'e map leading to results about their unitary 
representations, i.e. about eigenphases.
To transfer statistical properties of eigenphases of the QPM to 
energy eigenvalues, it
is necessary that the drift of eigenphases as a function of energy
be nearly parallel for all phases. Analytic evidence that this must 
be true for chaotic systems is given in 
\cite{Leyvraz&Seligman,Mendez%
%,Mendez2%
}, in agreement with the numerical results shown in 
Fig.~\ref{fig:Chaotique}.
To put this on a more quantitative basis we display
on Fig.~\ref{fig:Histo} histograms of the distributions
of the velocities of the eigenphases curves. The chaotic
case has a narrow distribution
while the integrable one shows long tails.
The present study is much more convincing in that respect
than another one for billiards using directly Bogomolny
theory\cite{Mendez}, which is garbled by the non unitarity
of the $T$ matrix for finite $\hbar$.

The difference seen in Fig.~\ref{fig:Histo} for the near 
integrable and the chaotic case is remarkable and it is thus tempting 
to consider the eigenphase velocity distribution as another
signature of classical chaos in quantum mechanics. While the narrow
distribution for chaotic systems is typical, the tails for the integrable
systems are not generic: they depend on the details of 
the partition of phase space by separatrices. 
In other examples \cite{Mendez} tails
have different shapes.

Summarizing: in this paper we proposed to interpret the multichannel quantum
defect theory as a quantum Poincar\'e map. This was implemented in detail in
the approximation that the absolute value of the electron angular momentum is
conserved, but it is quite clear that it is true in general. The new
interpretation allows for a more stable and efficient way to find solutions
to MQDT, particularly for near-degenerate levels. Beyond that, the 
approximate system has been used to exemplify quantum features of classically
chaotic systems. This can be now extended to use MQDT as a paradigm for QPM.
Indeed a study of the velocity distribution of eigenphases confirms
properties expected from other studies.

This work was partially supported by DGAPA (UNAM) project IN...
and the CONACYT grant 25192-E

%\bibliographystyle{physic}
%\bibliography{mqdtmap}

\end{document}